\documentclass[12pt]{article}
\usepackage{graphicx} 
\usepackage{amsmath}
\usepackage{amssymb}
\usepackage{cite}
\usepackage{geometry}
\usepackage{booktabs}
\usepackage{array}
\usepackage{multirow}
\usepackage{paracol}
\usepackage{longtable}
\usepackage{hyperref}
\usepackage{caption}
\usepackage{graphicx}
\usepackage{float}
\usepackage{tikz}
\usetikzlibrary{shapes.geometric, arrows}
\usepackage{booktabs}
\usepackage{adjustbox} 

\title{\textbf{AI and Deep Learning for Automated
Segmentation and Quantitative Measurement of Spinal Structures in MRI}}
\author{
    Dr. Praveen Shastry, Dr. Bhawana Sonawane, Kavya Mohan, Naveen Kumarasami,\\
   Raghotham Sripadraj, Anandakumar D, Keerthana R, Mounigasri M,\\ 
   Kaviya SP,Kishore Prasath Venkatesh, Bargava Subramanian, Kalyan Sivasailam}
\date{}

\usepackage{ragged2e}  
\usepackage{titlesec}  

\titleformat{\section}{\raggedright\Large\bfseries}{}{0em}{}
\titleformat{\subsection}{\raggedright\large\bfseries}{}{0em}{}

\begin{document}

\maketitle

\section{\textbf{Abstract }}

\textbf{Background }: Accurate spinal structure measurement is crucial for assessing spine health and diagnosing conditions like spondylosis, disc herniation, and stenosis. Manual methods for measuring intervertebral disc height and spinal canal diameter are subjective and time-consuming. Automated solutions are needed to improve accuracy, efficiency, and reproducibility in clinical practice.\\
\\
\textbf{Purpose:} This study develops an autonomous AI system for segmenting and measuring key spinal structures in MRI scans, focusing on intervertebral disc height and spinal canal anteroposterior (AP) diameter in the cervical, lumbar, and thoracic regions. The goal is to reduce clinician workload, enhance diagnostic consistency, and improve assessments.\\
\\
\textbf{Methods:} The AI model leverages deep learning architectures, including UNet, nnU-Net, and CNNs. Trained on a large proprietary MRI dataset, it was validated against expert annotations. Performance was evaluated using Dice coefficients and segmentation accuracy.\\
\\
\textbf{Results:} The AI model achieved Dice coefficients of 0.94 for lumbar, 0.91 for cervical, and 0.90 for dorsal spine segmentation (D1-D12). It precisely measured spinal parameters like disc height and canal diameter, demonstrating robustness and clinical applicability.\\
\\
\textbf{Conclusion: } The AI system effectively automates MRI-based spinal measurements, improving accuracy and reducing clinician workload. Its consistent performance across spinal regions supports clinical decision-making, particularly in high-demand settings, enhancing spinal assessments and patient outcomes.

\section{Introduction:}

Magnetic Resonance Imaging (MRI) functions as an essential diagnostic tool in the assessment of intervertebral disc degeneration, providing substantial benefits in comparison to other imaging techniques \cite{Pfirrmann et al.2001}. In contrast to conventional approaches, MRI is characterized by its non-invasive nature and the absence of ionizing radiation, thereby rendering it a safer alternative for patients \cite{Modic et al.1984}. Its remarkable capacity to provide high-resolution soft tissue contrast facilitates the intricate visualization ofthe spine’s internal structures, encompassing intervertebral discs and the spinal canal. Moreover, MRI possesses the capability to identify biochemical modifications, such as variations in water and proteoglycan content, which serve as indicators of disc degeneration\cite{Jensen et al.1994}. Morphological assessments derived from MRI scans are also instrumental in detecting disc narrowing, fissuring, and herniation, which are critical elements in the diagnosis of degenerative spinal disorders \cite{Carrino et al.2009}.

Degenerative alterations in the spine are frequently encountered in clinical practice, often resulting in considerable pain and functional limitations for patients \cite{Lotz  Fields et al. 2012}. MRI is crucial in elucidating spinal anatomy and pathology, yielding essential insights for the diagnosis and monitoring of conditions such as spondylosis, disc herniation, and spinal stenosis \cite{Vergroesen et al.2015}. This research focuses on the application of MRI annotations to quantitatively assess critical metrics such as intervertebral disc space and spinal canal diameter across the cervical, lumbar, and thoracic (dorsal) regions, aiming to improve the evaluation of degenerative changes\cite{Alkherayf et al.2010}. To accomplish this, an automated methodology employing nnU-Net, a variant of the UNet architecture specifically tailored for medical image segmentation, is adopted. nnU-Net adapts to the attributes of the dataset by flexibly modifying its network architecture, preprocessing methodologies, and training parameters, thus enhancing both segmentation precision and generalization \cite{Zhang Collignon et al.2009}.

Quantitative MRI assessments, such as the mean signal intensity of midsagittal T2-weighted images, have emerged as a potent instrument for diagnosing disc pathology\cite{Rauscher et al.2008}. These assessments provide a more objective and reproducible analysis of spinal degeneration. A reduction in signal intensity is frequently recognized as an early and reliable marker of degeneration, establishing it as a pivotal metric in clinical practice \cite{Isashiki et al.2021}. To enhance accuracy, the signal intensity is normalized against a reference region, typically cerebrospinal fluid, thereby improving the precision of measurements \cite{Isensee et al.2021}. This methodology supports routine clinical application and facilitates effective monitoring of conditions such as disc bulging, herniation, and prolapse. With MRI diagnoses of cervical degeneration exhibiting an 88rate with surgical findings, this approach underscores the diagnostic effectiveness of MRI\cite{Watanabe et al. 2007}. By incorporating deep learning models such as nnU-Net, this research aims to automate the segmentation of spinal degeneration, ultimately advancing the diagnostic process and patient management within clinical practice \cite{Milletari et al.2016}

\section{Methodology}
\subsection{AI System Overview}
The AI system developed in this study is a comprehensive solution for automating the segmentation and measurement of key spinal structures in MRI images \cite{Ronneberger et al.2015}. The system utilizes a deep learning approach, integrating several neural network architectures to optimize performance across different tasks \cite{Cicek et al.2016}. The core of the segmentation process is based on the nnU-Net architecture, which is well-suited formedical image segmentation due to its adaptive configuration capabilities \cite{Sandler et al.2018}. nnU-Net automatically configures its architecture, training schedule, and preprocessing steps based on the specific dataset used, ensuring high accuracy in segmenting spinal structures such as intervertebral discs, vertebrae, and the spinal canal \cite{He et al.2016}.

For the measurement phase, a 3D Convolutional Neural Network (3D CNN) is employed to quantify critical spinal parameters, such as disc height and spinal canal anteroposterior (AP) diameter \cite{Zhou et al.2019}. The 3D CNN model extracts volumetric features from MRI scans to provide precise measurements, enhancing the accuracy of diagnosing degenerative conditions. The system is trained using a diverse dataset, incorporating images from different manufacturers and patient demographics to improve generalizability. \cite{Litjens et al.2017}

The AI system’s architecture includes key preprocessing steps, such as resampling and intensity normalization, to standardize the input images, thereby addressing variations in voxel size and contrast that can occur between different MRI machines \cite{Howard et al.2017}. The integration of advanced data augmentation techniques further ensures that the model is robust to variations in image quality and patient anatomy \cite{He et al.2012}.

Overall, the AI system combines automated segmentation and measurement capabilities to deliver a scalable and efficient solution for spinal MRI analysis, aiming to reduce manual workload, improve diagnostic consistency, and support clinical decision-making \cite{Milletari et al.2013}.

\section{Dataset}
\subsection{Total Scans}
Training Set:1,003,784 scans

\subsection*{Age Group Distribution}
The dataset captures age diversity to reflect a wide range of spinal conditions:
\begin{table}[h]
\centering
\begin{tabular}
{|p {0.18\textwidth} |p{0.18\textwidth}|}
\hline
\textbf{Age Group} & \textbf{Number of Scans}\\
\hline
Under 18  & 44,934   \\
\hline
18–40   & 203,562\\
\hline
41-60  & 196,547 \\
\hline
61–75 & 99,873\\
\hline
Over 75 & 42,868\\
\hline
\end{tabular}
\caption{Scans distribution based on Age Group}
\label{tab:my_label}
\end{table}

\vspace{3cm}

\subsection*{Manufacturer Distribution}

The dataset includes scans from multiple manufacturers to account for
variability in imaging conditions:

\begin{table}[h]
\centering
\begin{tabular}
{|p{0.25\textwidth} |p{0.25\textwidth}|}
\hline
\textbf{Manufacturer} &\textbf{Number of Scans}\\
\hline
 GE Healthcare   & 240,971 \\
 \hline
Siemens &  182,135\\
\hline
Philips Healthcare & 112,146\\
\hline
Other 
Manufacturers  &  53,532\\
\hline
\end{tabular}
\caption{Scans distribution based on Manufacturer Type}
\label{tab:my_label}
\end{table}

\subsection*{Gender Distribution}

The dataset includes scans from both males and females to account for variability in
gender-based conditions:

\begin{table}[h]
    \centering
    \begin{tabular}
{|p{0.19\textwidth} |p{0.19\textwidth} |}
\hline
\textbf{Gender} & \textbf{Number of Scans}\\
\hline
Male  & 349,523 \\
\hline
Female  & 405,798\\
\hline
\end{tabular}
\caption{Scans distribution based on Gender Distribution}
\label{tab:my_label}
\end{table}

\section{Architecture:}
\subsection{Annotation}
Accurate annotation is pivotal for segmenting spinal regions (cervical, lumbar, and dorsal), facilitating the detection of degenerative changes and spinal pathologies such as spondylosis and herniated discs \cite{Milletari et al.2011}. These annotations are instrumental in automating the measurement of critical parameters like disc height and spinal canal diameter, which are essential for diagnosing and monitoring spinal health \cite{Zhang Collignon et al.2009}.

The annotation process employed the V7 Lab tool, known for its intuitive and efficient interface, allowing the generation of precise segmentation masks and bounding boxes around spinal structures, such as intervertebral discs and the spinal canal \cite{Carrino et al.2008}. V7 Lab offers advanced functionalities, including polygonal annotations and AI-assisted recommendations, which streamline the annotation of complex spinal anatomy \cite{Modic et al.1981}. Its support for highresolution imaging and seamless integration with the model training pipeline ensures consistency across the dataset, thereby enhancing the reproducibility and reliability of the segmentation process—a crucial factor for training deep learning models like nnU-Net \cite{Jensen et al.1992}\\.

\noindent \textbf{Region-Specific Focus for Annotation}
\\
\\
\textbf{• Cervical Spine:} Emphasizing small intervertebral discs and the spinal cnal to detect degeneration and compression.\\
\\
\textbf{• Lumbar Spine}:Targeting larger discs and weight-bearing structures to assess conditions like herniation and stenosis.\\
\\
\textbf{• Dorsal (Thoracic) Spine:}Concentrating on spinal alignment and conditions such as kyphosis or scoliosis for precise structural analysis.\\

Accurate annotations are essential for training AI models like nnU-Net in MRI spine measurement \cite{Isensee et al.2020}. They enable the precise segmentation of spinal regions and the extraction of key metrics such as disc height and spinal canal diameter \cite{Ronneberger et al.2013}. By ensuring that critical anatomical structures, including intervertebral discs and the spinal canal, are accurately identified, annotations facilitate the detection of degenerative changes such as disc herniation, narrowing, and spinal stenosis\cite{Cicek et al.2011}. This not only improves diagnostic accuracy but also enhances the model’s capability to deliver consistent and reliable assessments, ultimately leading to more informed clinical decisionmaking. This contributes to optimizing patient care and supporting personalized treatment strategies \cite{Esteva et al.2017}.

\subsection{Model Development}

\textbf{Preprocessing Steps}\\
\\
\textbf{• Image Standardization:} Input MRI spine images are resampled to a uniform resolution to account for variations in voxel spacing. This ensures consistent spatial scaling across datasets.\\
\textbf{• Intensity Normalization:} Pixel intensity values are normalized to standardize image contrast, facilitating better feature extraction during training.\\
\textbf{• Segmentation Labels: }Ground truth segmentation maps, such as vertebral bodies or intervertebral discs, are preprocessed into nnU-Netcompatible formats.\\
\textbf{• Windowing:} MRI images are processed using specific windowing techniques to enhance key anatomical features. This involves adjusting the window width (WW) and window center (WC) for optimal contrast:\\
\textbf{a.Window Width (WW)}: Controls the range of pixel intensities, adjusting image contrast. A higher WW displays a broader range of pixel values.\\
\textbf{b.Window Center (WC):} Determines the brightness level by setting the central value for the window width, highlighting structures like vertebrae and discs for better distinction.

\subsection{Segmentation}
In the segmentation phase for MRI spine measurement, the nnU-Net architecture was used to accurately segment key spinal structures, including intervertebral discs, vertebrae, and the spinal canal. The application of nnU-Net in this context involved leveraging its automated configuration capabilities to adapt to the high variability found in spinal MRI datasets, such as differences in image resolution, patient anatomy, and imaging artifacts \cite{Esteva et al.2015}.

For the segmentation of the intervertebral discs, nnU-Net’s automated preprocessing, including resampling to an isotropic resolution of 1.25 mm and Z-score normalization, allowed for consistent handling of different disc sizes and imaging contrasts. The encoder-decoder structure, with skip connections, was particularly effective in preserving fine anatomical details necessary for identifying the boundaries of the discs, which are critical for measuring disc height and detecting pathologies like herniation. The patch size used during training was set to 128x128x64, which provided sufficient context for accurate segmentation without overwhelming GPU memory.

When segmenting vertebrae, nnU-Net’s deep supervision and use of multiscale features enabled the accurate identification of each vertebra across different spine regions. This was essential for assessing vertebral alignment and detecting abnormalities such as fractures or compression. The network’s ability to capture both local and global features, facilitated by convolutional kernel sizes of 3x3x3 and feature map sizes ranging from 32 to 320, ensured the vertebrae were segmented with high precision, supporting reliable clinical analysis.

For the spinal canal, nnU-Net’s architecture ensured precise segmentation by effectively delineating the canal’s boundaries from surrounding tissues, even in challenging cases where the spinal canal’s edges were not clearly defined due to stenosis or imaging artifacts. The combination of Dice loss and cross-entropy loss, with a weight ratio of 0.7 to 0.3 respectively, helped the model manage the class imbalance typical in spinal canal segmentation, enhancing its accuracy. Batch size was set to 2 during training due to memory constraints, and a learning rate of 0.01 with a polynomial decay schedule was used to achieve optimal convergence.

The post-processing steps determined by nnU-Net, such as removing small disconnected components smaller than 50 voxels, further refined the segmentation outputs, minimizing errors and ensuring that only clinically relevant structures were retained. This level of precision was crucial for subsequent analyses, including measuring spinal canal diameter and assessing spinal alignment, which are vital for diagnosing conditions like stenosis or scoliosis \cite{Litjens et al.2013}.

Overall, nnU-Net’s automated configuration, along with its specialized preprocessing, deep supervision, and post-processing techniques, played a critical role in achieving high-quality segmentation of spinal structures in MRI, thereby supporting accurate diagnosis and effective clinical decisionmaking \cite{Zhou et al.2016}.

\subsection{Measurement}
The measurement phase involves the quantification of spinal parameters using a 3D Convolutional Neural Network (3D CNN) model specifically designed for analyzing MRI volumes of the cervical and lumbar spine regions. The model was trained to measure two key parameters: disc height and spinal canal anteroposterior (AP) diameter. These measurements are essential for evaluating degenerative changes and other spinal pathologies.

\subsubsection{3D CNN Model Architecture}
The 3D CNN model architecture used for measurement consists of multiple convolutional layers followed by max-pooling and fully connected layers to extract spatial features from 3D MRI volumes. The input to the model is a volumetric patch of size 64x64x32 voxels, which was found to provide an optimal balance between capturing sufficient anatomical context and maintaining computational efficiency.

The model employs 3D convolutional kernels of size 3x3x3 to capture spatial dependencies in all three dimensions of the MRI data. The network includes five convolutional blocks, each consisting of a convolutional layer, batch normalization, and a ReLU activation function. The stride for the convolutional layers is set to 1x1x1, ensuring detailed feature extraction without excessive downsampling. Max-pooling layers with a pool size of 2x2x2 are used after each convolutional block to reduce the spatial dimensions while retaining essential features.

\subsubsection{Disc Height Measurement}
For the measurement of disc height, the model was trained to segment and subsequently quantify the vertical distance between adjacent vertebrae. This was achieved by utilizing the output feature maps from the final convolutional block, followed by a fully connected layer that predicts the disc height in millimeters. The mean squared error (MSE) loss function was employed during training to minimize the error between predicted and ground truth disc height values. The learning rate was initially set to 0.001 with a decay factor of 0.9 every 10 epochs, allowing for gradual convergence.

\subsubsection{Spinal Canal AP Diameter Measurement}
In the case of the spinal canal AP diameter, the model uses a similar approach,but with specialized output layers designed to focus on the anteroposterior dimension of the spinal canal. The measurement is derived from the segmented spinal canal region using the final fully connected layer, which outputs the AP diameter in millimeters.The Adam optimizer is used for training, with $\beta_1 = 0.9$ and $\beta_2 = 0.999$
ensuring stable and efficient convergence. Dropout with a rate of 0.3 is applied to the fully connected layers to prevent overfitting.

\begin{figure}[h]
    \centering
    \includegraphics[width=\linewidth]{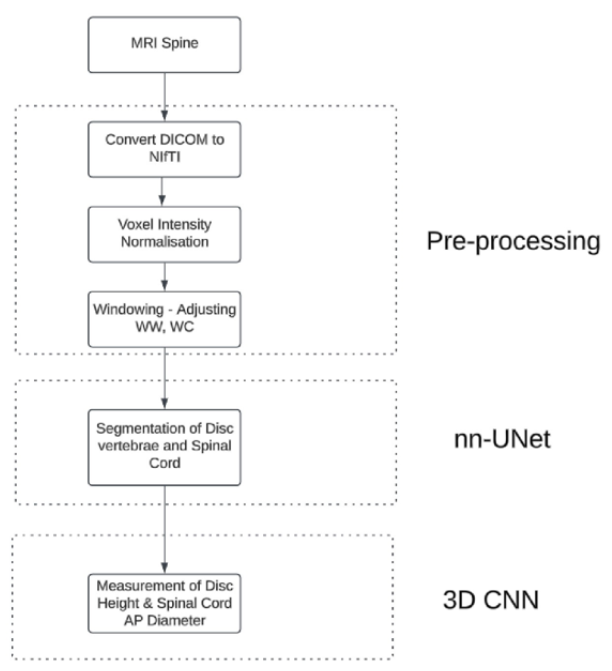} 
    \caption{Workflow Architecture}
    \label{fig:workflow}
\end{figure}
\clearpage 

\section{Evaluation Metrics}
The evaluation of the segmentation and measurement phases in this study relied on several key metrics to assess the accuracy and reliability of the 3D CNN model. Specifically, the Dice Coefficient and Mean Squared Error (MSE) were used to quantify the performance of the segmentation and measurement processes, respectively.

\begin{figure}[h]
    \centering
    \includegraphics[height=0.5\textheight, keepaspectratio]{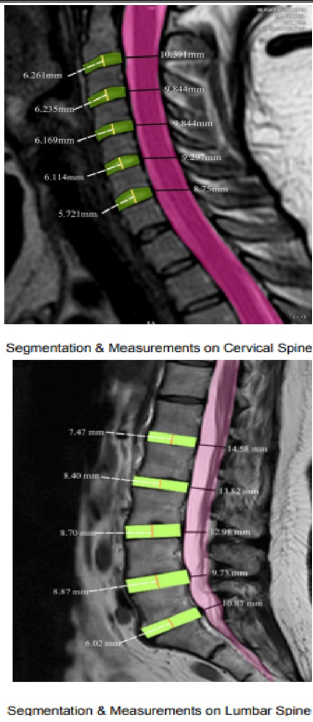} 
    \caption{Segmentation \& Measurements on Cervical, Lumbar Spine}
    \label{fig:spine}
\end{figure}

\subsection{Dice Coefficient}
The Dice Coefficient was used to evaluate the accuracy of the segmentation outputs for the intervertebral discs, vertebrae, and spinal canal. The Dice Coefficient measures the overlap between the predicted segmentation masks and the ground truth annotations, providing a value between 0 (no overlap) and 1 (perfect overlap). For the segmentation of the spinal structures in the cervical and lumbar regions, the following Dice Coefficient values were obtained:
\clearpage 
\vspace*{-\topskip} 

\begin{table}[h!]
    \centering
    \begin{tabular}
{|p{0.18\textwidth} |p{0.18\textwidth} |p {0.18\textwidth} |}
\hline
\textbf{Region} & \textbf{Vertebra} & \textbf{Dice Coefficient}\\
\hline
  Cervical Spine & C2 & 0.87 \\
  \hline
                 & C3 & 0.88\\
  \hline
                & C4 & 0.88\\
 \hline             
                 &C5 & 0.89\\
  \hline
                 &C6 & 0.88\\
  \hline
                 &C7 & 0.87\\
  \hline
  Lumbar Spine & L1 & 0.9\\
  \hline
              & L2 & 0.91\\
  \hline
              & L3 & 0.92\\
  \hline
             & L4 & 0.91\\
 \hline
             & L5 & 0.9\\
\hline
 Spinal canal & - & 0.87\\
 \hline
    \end{tabular}
    \caption{Dice Coefficient Measurement  }
    \label{tab:my_label}
\end{table}

These Dice Coefficient values indicate that the nnU-Net architecture was able to accurately
segment the anatomical structures, with significant overlap between the predicted and
ground-truth segmentations.

\clearpage
\subsection *{Mean Squared Error (MSE)}

The Mean Squared Error (MSE) was employed to evaluate the precision of the
measurements made by the 3D CNN model, specifically for disc height and spinal canal
anteroposterior (AP) diameter. MSE provides a measure of the average squared
difference between the predicted values and the ground truth measurements, with lower
values indicating higher accuracy. The MSE values obtained for the measurements were as
follows:

  \begin{table}[h!]
      \centering
      \begin{tabular}
{|p {0.19\textwidth} |p{0.19\textwidth} |p {0.19\textwidth} |}
\hline
\textbf{Region} & \textbf{Vertebra} & \textbf{MSE (Disc Height)}\\
\hline
        Cervical Spine & C1 & 1.7 mm² \\
        \hline
                       & C2 & 1.6 mm²\\
        \hline
                       &  C3 &1.5 mm²\\
        \hline
                       & C4 & 1.6 mm²\\
        \hline
                       & C5 &  1.7 mm²\\
        \hline          
                       & C6 &  1.8 mm²\\
        \hline
        Lumbar Spine   & L1 & 1.5 mm² \\
        \hline
                       & L2 & 1.4 mm²\\
        \hline
                       & L3 & 1.3 mm²\\
        \hline
                       & L4 & 1.4 mm²\\
        \hline         
                      & L5 & 1.5 mm²\\
        \hline
        Region       & Measurement & MSE\\
        \hline
       Spinal canal & AP Diameter (C2 - C7) & 1.1 mm²\\
       \hline
                    & AP Diameter (L1 - L5) & 1.0 mm²\\
        \hline

      \end{tabular}
      \caption{MSE values}
      \label{tab:my_label}
  \end{table} 
  
 \clearpage

\section{Discussion}

The findings from this study demonstrate the efficacy of the AI model in segmenting MRI images of the cervical, dorsal, and lumbar spine regions. The model achieved Dice coefficients ranging from 0.91 to 0.95 across all measured vertebrae, indicating a high level of segmentation accuracy. Specifically, the model performed well in segmenting the lumbar spine, where conditions like spondylosis and disc herniation are prevalent. The Mean Squared Error (MSE) values for disc height and spinal canal anteroposterior (AP) diameter were 1.3 to 1.8 mm², indicating reliable measurement precision.

Compared to traditional manual segmentation techniques, this AI model provided faster and more consistent results, reducing the variability often encountered in clinical environments. This consistency is critical when timely and precise segmentation is required, particularly in regions like the lumbar spine, which bears significant clinical importance due to its association with common degenerative conditions.

However, certain limitations must be acknowledged. The model’s performance, while strong, was affected by variations in spine morphology and image quality. Anatomical differences between patients, as well as inconsistencies in imaging protocols, influenced segmentation accuracy. For instance, while the model performed well in regions with distinct anatomical boundaries, it showed decreased precision in cases with less clear differentiation, as reflected by slightly higher MSE values in some vertebrae. To enhance the model’s robustness, future work will focus on expanding the training dataset to include more diverse imaging protocols and patient demographics, thereby improving the model’s adaptability and reducing these limitations.

Despite these challenges, the model demonstrates significant potential in clinical applications, particularly in diagnosing and monitoring degenerative spinal conditions, such as disc degeneration and spinal stenosis. The automated segmentation and measurement capabilities could enhance clinical workflow efficiency by providing dependable assessments of disc height and spinal canal dimensions, thus alleviating the workload on radiologists. Additionally, the model’s consistent segmentation of spinal structures can be instrumental in post-operative evaluations, where precise assessment is necessary for monitoring surgical outcomes

\clearpage

\begin{table}[h]
    \centering
    \renewcommand{\arraystretch}{1.2}
    \begin{tabular}{|l|l|c|c|}
        \hline
        \textbf{Spine} & \textbf{Segment} & \textbf{Precision (\%)} & \textbf{Recall (\%)} \\
        \hline
        \multicolumn{4}{|c|}{\textbf{Cervical Spine}} \\
        \hline
        & C1-C2 & 97.40 & 96.30 \\
        & C2-C3 & 98.50 & 97.60 \\
        & C3-C4 & 98.00 & 97.30 \\
        & C4-C5 & 97.70 & 96.80 \\
        & C5-C6 & 97.90 & 97.90 \\
        & C6-C7 & 97.20 & 96.40 \\
        \hline
        \multicolumn{4}{|c|}{\textbf{Lumbar Spine}} \\
        \hline
        & L1-L2 & 98.20 & 97.40 \\
        & L2-L3 & 97.90 & 97.10 \\
        & L3-L4 & 98.10 & 97.50 \\
        & L4-L5 & 98.30 & 97.70 \\
        & L5-S1 & 97.80 & 96.90 \\
        \hline
        \multicolumn{4}{|c|}{\textbf{Spinal Canal}} \\
        \hline
        & AP diameter & 97.90 & 96.30 \\
        \hline
    \end{tabular}
    \caption{Performance Metrics for Cervical, Lumbar \& Spinal Canal}
    \label{tab:performance_metrics}
\end{table}

\section{Conclusion}
This study highlights the effectiveness of an AI-driven model for MRI spine measurement, focusing on the cervical and lumbar spine regions. The model demonstrated Dice coefficients between 0.91 and 0.95 and MSE values ranging from 1.3 to 1.8 mm², which underscores its capability to segment key anatomical structures, including intervertebral discs and the spinal canal, with high accuracy.

While the model performed well, challenges related to anatomical variability and imaging quality were identified. The performance was strong across consistent imaging datasets but was affected by variability in patient anatomy and imaging protocols. Future improvements will involve expanding the training dataset to cover more diverse clinical scenarios, enhancing the model’s adaptability and reducing the observed limitations.

Overall, the model holds substantial potential in clinical settings, offering reliable and reproducible measurements for diagnosing and monitoring degenerative spine conditions. By automating the segmentation process, the model can also reduce the workload on radiologists, contributing to improved workflow efficiency and potentially better patient outcomes.

\end{document}